\documentclass{JINST}

\title{The Data Acquisition System for LZ}

\author{Eryk Druszkiewicz$^a$\thanks{Corresponding author.}~, for the LZ Collaboration
\\
\llap{$^a$}Department of Physics and Astronomy, University of Rochester, Rochester, New York 14627, USA\\

E-mail: \email{eryk.druszkiewicz@rochester.edu}}

\abstract{The	Data Acquisition System	of	the	LZ	experiment,	the	10-tonne	dark	matter	detector	to	be	installed	at	the	
Sanford	Underground	Research	Facility	(SURF),	will	collect	signals	from	788	
photomultiplier	tubes	(PMTs).	Because	the	signals	from	the	time	projection	
chamber	PMTs	will	be	passed	through	dual-gain	amplifiers,	the	DAQ	system	will	
collect	waveforms	from	a	total	of	1276	channels,	using	custom	built,	32-channel,	
FPGA-based	digital	signal	processors.	The	appropriately	conditioned	signals	will	be	
digitized	at	100	MHz	with	14-bit	resolution.	Based	on	actual	measurements	with	a	
small-scale	prototype	system,	the	LZ	DAQ	is	expected	to be	able	to	handle	a	
maximum	sparsified	data	rate	of	~1500	MB/s.	During	calibrations,	it	is	estimated	
that	only	33\%	of	the	system	resources	are	utilized.	 The	digital	filters	that	are	used	
for	data	selection	operate	with	an	aggregate	throughput	in	excess	of	595,000~MB/s.	 Data	selection	decisions	are	based	on, for	example,	the	amount	of scintillation	(S1)	and	photoluminescence S2 light,	S1	and	S2	hit-patterns,	and	total	
energy	deposition.}


\begin{document}

\section{Introduction}\label{sec:intro}
\begin{figure}[b] 
\centering
\includegraphics[width=1.0\textwidth]{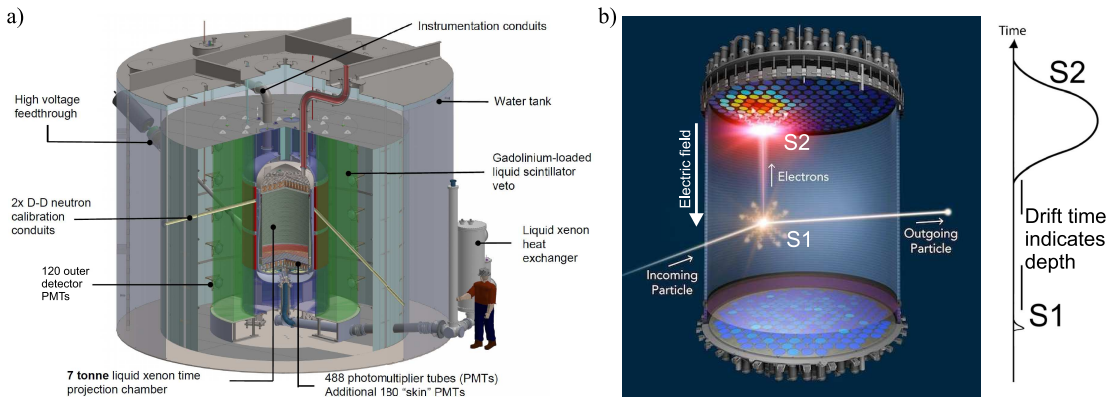}
\caption{a) 3D model of the LZ detector with the main parts indicated. b) Diagram depicting the principle of event detection in LZ.}
\label{fig:LZEventOperation}
\end{figure}

\indent LZ is the next-generation Dark Matter Search experiment that will be deployed at the Sanford Underground Research Facility (SURF) as a successor to the Large Underground Xenon (LUX) detector~\cite{LUXExp}. LZ is dual-phase xenon detector that will have a total xenon mass of 10 tonne. The TPC region accounts for 7 tonne with a fiducial volume of 5.6 tonne~\cite{LZCDR}. Figure~\ref{fig:LZEventOperation}a depicts the main parts of the detector. The detector will be instrumented with 488 TPC PMTs, 180 skin PMTs (looking at the outer TPC Xe volume), and 120 outer detector PMTs. 
\newline
\indent The principle of event detection is shown in Fig.~\ref{fig:LZEventOperation}b. 
When an incoming particle interacts with a xenon atom scintillation photons and ionization electrons are created. The scintilation photons result in a prompt S1 signal. The electrons drift towards the liquid surface and are extracted into the gas phase where a secondary wider S2 signal is generated. From the relation between the S1 and S2 signals, the type of interaction can be inferred. The time separation of the S1 and S2 signals provides information about the depth of the initial interaction, while the S2 illumination pattern on the top PMT array can be used to determine the planar position of the initial interaction~\cite{LUXExp}.

\newpage

\section{Data Acquisition System}\label{sec:daq}

\begin{figure}[b] 
\centering
\includegraphics[width=1.0\textwidth]{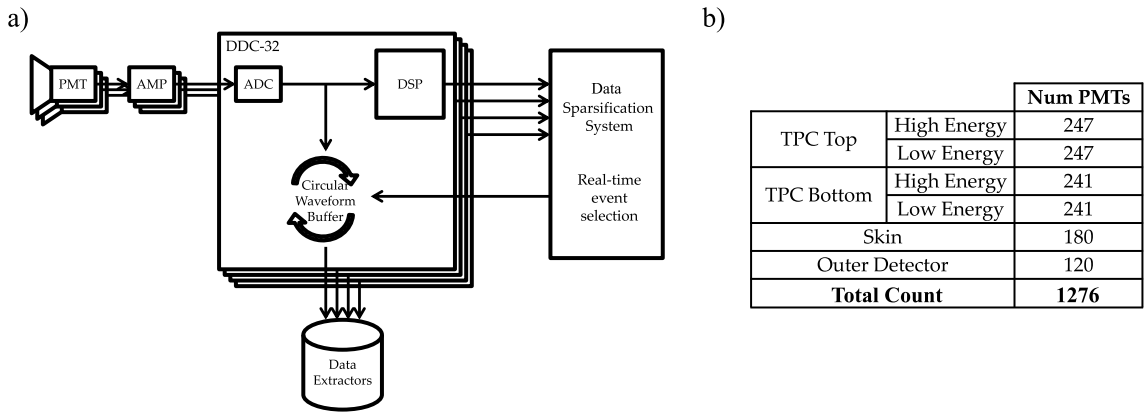}
\caption{a) Simplified depiction of how PMT signals are going to be collected. b) PMT channel count breakdown in the LZ experiment.}
\label{fig:NumChannelDAQOperation}
\end{figure}

\indent A simplified schematic operation of the LZ DAQ system is shown in Fig.~\ref{fig:NumChannelDAQOperation}a. At the xenon to air interface, the signals are shaped and amplified with an array of custom built amplifiers. The signals are digitized and sent to the internal circular waveform buffers where baseline-suppressed pulse waveforms are stored~\cite{LUXDAQ}. 
Reduced quantities, such as hit-vectors, multiplicity counts, and pulse area, are extracted and passed to the data sparsification system (DS), where  they are analyzed in real-time in search for events of interest. Once a potentially good event is identified, the data are off-loaded from the circular waveform buffers for further off-line analysis. 
\newline
\indent The breakout amplifiers, provide dual gain outputs for the TPC PMT signals and single gain outputs for the skin and outer detector PMT signals. The dual-gain maximizes the available dynamic range and thus extends the spectrum of interactions that can be probed with the LZ detector \cite{JunProceeding}. 
The PMT channel count is summarized in Fig.~\ref{fig:NumChannelDAQOperation}b. Thus the LZ DAQ system must be able to handle at least 1276 digitization channels.
\newline
\indent The digitization of the analog signals is performed at 100	MHz	with	14-bit	resolution, using custom built,	32-channel,	FPGA-based DDC-32 digital	signal	processors. Based on the LUX event sizes, it has been determined that the amount of internal FPGA memory is sufficient to hold events expected in LZ. For example, a typical ${\rm {}^{83m}Kr}$ event is predicted to occupy about 5\% of the available waveform memory. The hardware is developed in collaboration with Skutek Instrumentation~\cite{Skutek}, while the entire firmware and software is developed by the LZ collaboration. The decision to use custom hardware was based on cost savings and previous experience from the LUX experiment that showed that full access to hardware design details as well as control over the software and firmware is crucial to the success of experiments of this size and complexity.

\begin{figure}[b] 
\centering
\includegraphics[width=1.0\textwidth]{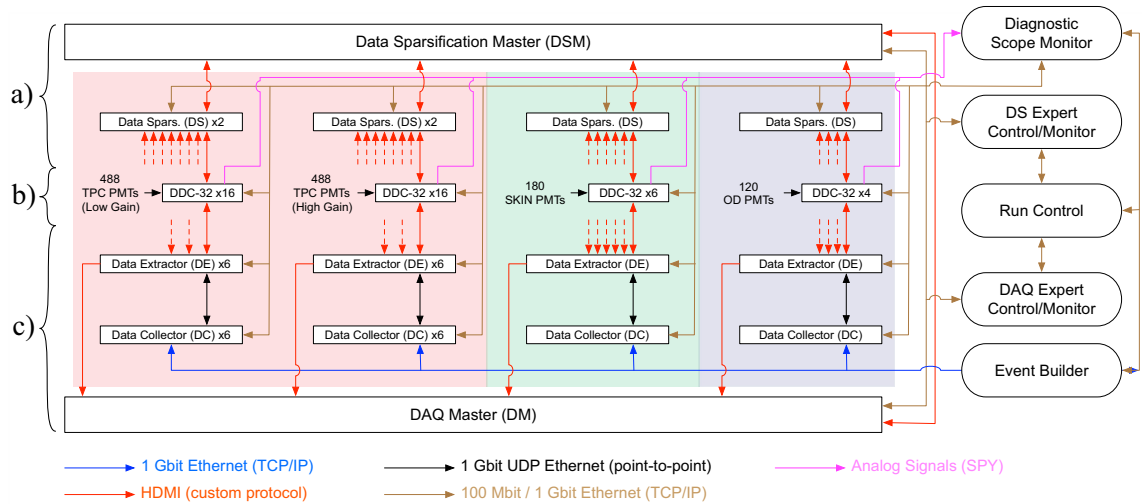}
\caption{Schematic top-level view of the LZ Data Acquisition System. a) Data Sparsification side of the system. b) Common set of digitizers shared between the Data Collection and Sparsification sides of the system. c) Data Collection side of the system.}
\label{fig:LZDAQ_Schematic}
\end{figure}

\indent Figure~\ref{fig:LZDAQ_Schematic} schematically shows the top-level representation of the entire LZ DAQ system. The data collection (Fig.~\ref{fig:LZDAQ_Schematic}c) and data sparsification sides of the system share a common set of digitizers (Fig.~\ref{fig:LZDAQ_Schematic}b). This approach is beneficial because both sides of the system work with a single common digital representation of the incoming PMT signals. This greatly simplifies off-line cross checking and system performance evaluation.
\newline
\indent Most of the data is sent between system boards using custom serial links, built on FPGA SerDes blocks~\cite{XilinxSerDes}, using HDMI cables as the physical transportation medium (indicated with red arrows in Fig.~\ref{fig:LZDAQ_Schematic}). In our prototype system~\cite{DevProceeding}, the links have been made to work reliably at 225~MB/s. The HDMI cables will also be used to distribute the global clock and synchronization signals from the DAQ and Sparsification Master units to the lower level units.
\newline
\indent When an event of interest is off-loaded, it is first extracted by the Data Extractors (DE) where the serial waveform data is packaged into a series of FPGA generated Ethernet packets. These packets are sent to Data Collectors (DC). The dedicated Ethernet link between each DE and DC pair is a 1~gigabit User Datagram Protocol (UDP) link. Because we implement consistency checking at the application level and this UDP link is setup in a point-to-point configuration, it can be used reliably, with minimal overhead and close to full gigabit bandwidth. In our prototype system, with some tweaking of the DC network stack, we have been able to send data continuously at 109~MB/s without corruption or data loss. This will be the most stressed data off-loading link in the DAQ system and yet we do not expect to utilize more than 33\% of its bandwidth. Considering that there are fourteen DE/DC pairs in the system, we expect it to be able to collect data at a peak rate of 1500~MB/s. 
The data stored on the DCs is organized by channels and made available to the Event Builder (EB) for full event assembling. 
\newline
\indent The DCs are server-grade, rack-mountable (2U) workstations, which are custom built from off-the-shelf parts. We have assembled a few units and verified reliable data transfer form the DE to the solid-state drive (SSD) in the DC. The DCs will have adequate storage resources (\textasciitilde1~week) to buffer the collected data in case of network outage problems, which might temporarily prevent the data from being transferred to off-site storage facilities. 
\newline
\indent Each board in this system has an on-board Blackfin processor~\cite{AnalogBF} which runs a $\mu$Clinux operating system~\cite{uClinux}. This allows for convenient system setup and operation over a dedicated Ethernet control network (indicated with brown arrows in Fig.~\ref{fig:LZDAQ_Schematic}).

\newpage

Because of the sheer volume of data that come from the PMTs, where just the volume associated with PMT dark counts amounts to 8.5~PB over the course of the experiment, a real-time event selection is needed and is done by the data sparsification side of the system (Fig.~\ref{fig:LZDAQ_Schematic}a).  
\newline
\indent All digitization channels have a set of two digital filters that provide an output that is proportional to the area of the input pulse and perform real-time baseline subtraction. One filter has a width that is matched to a typical S1 pulse contribution seen by a single PMT (\textasciitilde60~ns FWTM) while the other is matched to a typical S2 pulse with a FWTM width of a few microseconds. The output of these filters is used to create hit and multiplicity vectors. Based on the experience of signal processing for the LUX experiment \cite{ErykTrigPaper}, as the starting point, we plan to use filters shown in Fig. 4. Since these are essentially integrating filters, a large S1 pulse may trigger the S2 filter if its area exceeds the S2 filter threshold. We are working on techniques based on the relative outputs of the S1 and S2 filters that allow for more robust S1/S2 pulse type discrimination. The total expected throughput of the data processed by these filters is estimated to exceed 595,000~MB/s~\cite{ErykThesis}.
\newline
\indent Each DDC-32 digitizer provides a digital sum of all or selected channels, allowing for construction of a total sum at the Data Sprasification Master (DSM)(Fig.~\ref{fig:LZDAQ_Schematic}a). This sum is processed by its own set of digital filters and allows the total pulse area to be part of the decision making process.
\newline
\indent DS information, such as hit/multiplicity vectors and the sum waveform are off-loaded with each event and merged with the DAQ data stream. This enables off-line data sparsification performance verification and optimization.
\newline
\indent The entire system is intentionally grouped by main detector sections (TPC, skin, and outer detector). Such grouping allows for more optimal data processing, because each of the groups has different event characteristics (e.g. no S2-like signals in the outer detector). This simplifies event selection for each group. The system will accommodate external triggers, such as a system wide heartbeat or a LED calibration synchronization signals.
\newline
\indent The LZ DAQ System is designed to handle event rates up to \textasciitilde250kHz. This is much higher than the highest expected event rates in the LZ detector, which will occur during LED calibrations (\textasciitilde4 kHz).
\begin{figure}[b] 
\centering
\includegraphics[width=1.0\textwidth]{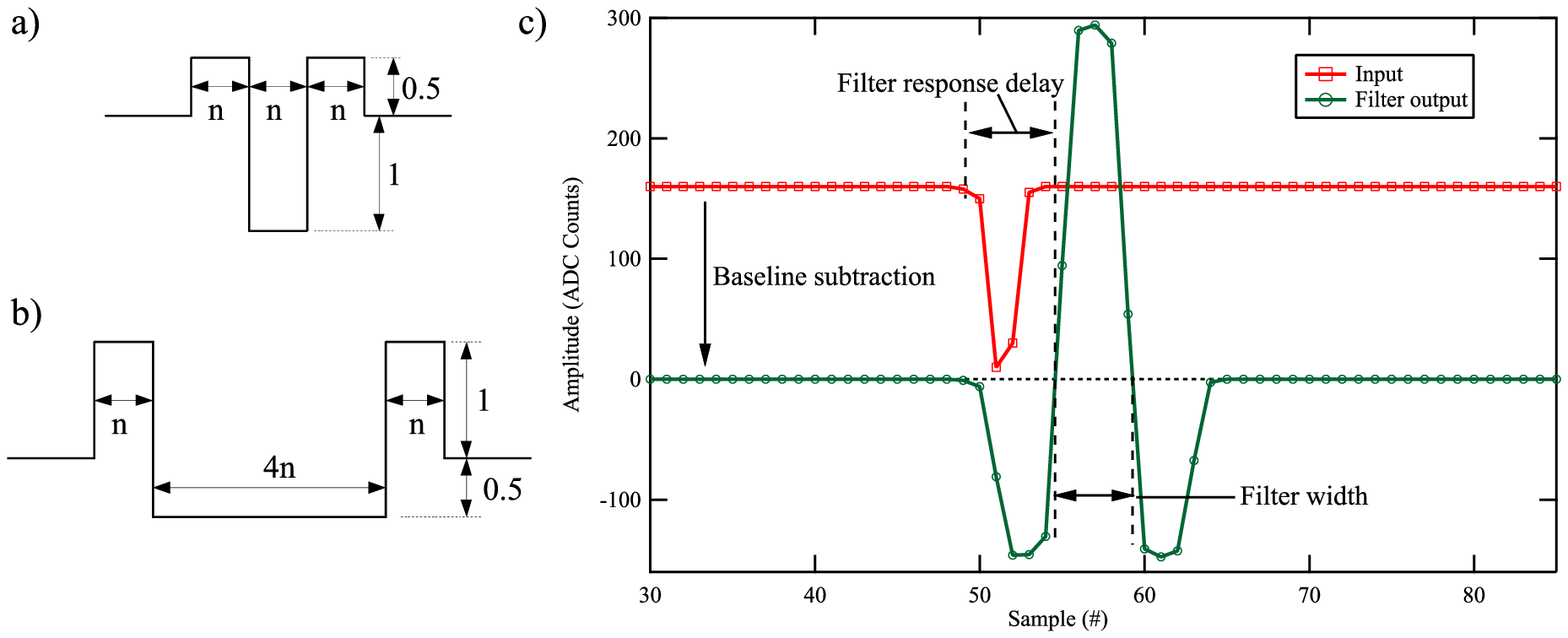}
\caption{a) S1 filter whose width (n) is matched to typical S1-like pulse. b) S2 filter whose width (4n) is matched to typical S2-like pulse. c) Example of S1 filter in operation.}
\label{fig:FiltersFigure}
\end{figure}

\newpage

\section{Conclusions}\label{sec:conclusions}
The LZ data acquisition system must be capable of digitizing and processing over 1200 PMT channels. 
The system under development provides real-time event selection and a total data off-loading rate capability of \textasciitilde1500~MB/s. 
The system utilizes computational parallelism offered by FPGA technology and processes waveforms at rates in excess of 595,000~MB/s.
A test system, representative of two full signal chains, is currently being put together. This test system will allow for verification of signal propagation from actual PMTs, all the way to event files on disk. Additionally elements such as data throughput, system control and monitoring will be developed and optimized. This first test stage is expected to be completed by spring of 2016 \cite{DevProceeding}.

\acknowledgments

The work was partially supported by the U.S. Department of Energy (DOE) under award numbers DE- SC0012704, DE-SC0010010, DE-AC02-05CH11231, DE-SC0012161, DE-SC0014223, DE-FG02- 13ER42020, DE-FG02-91ER40674, DE-NA0000979, DE-SC0011702, DE-SC0006572, DE-SC0012034, DE-SC0006605, and DE-FG02-10ER46709; by the U.S. National Science Foundation (NSF) under award numbers NSF PHY-110447, NSF PHY-1506068, NSF PHY-1312561, and NSF PHY-1406943; by the U.K. Science \& Technology Facilities Council under award numbers ST/K006428/1, ST/M003655/1, ST/M003981/1, ST/M003744/1, ST/M003639/1,\linebreak ST/M003604/1, and ST/M003469/1; and by the Portuguese Foundation for Science and Technology (FCT) under award numbers CERN/FP/123610/2011 and PTDC/FIS-NUC/1525/2014.

\end{document}